\documentclass[runningheads]{llncs}
\usepackage{amssymb}
\usepackage{amsmath}
\usepackage[utf8]{inputenc}
\usepackage[english]{babel}
\usepackage{graphicx}
\usepackage{listings}

\begin{document}
\title{Commutative Event Sourcing vs. Triple Graph Grammars}
\author{Sebastian Copei \and
Albert Z\"undorf}
\authorrunning{S. Copei, A. Z\"undorf}
\institute{Kassel University, Germany}
\maketitle              % typeset the header of the contribution
\begin{abstract}
This paper proposes Commutative Event Sourcing as a 
simple and reliable mechanism for model synchronisation,
bidirectional model to model transformations, incremental updates, 
and collaborative editing. 
Commutative Event Sourcing is a restricted form of a 
Triple Graph Grammar where the rules or editing commands 
are either overwriting or commutative. This 
restriction gets rid of a lot of Triple Graph Grammar 
complexity and it becomes possible to implement 
model synchronisation manually. Thus, you are not  
restricted to Java as your programming language and 
you do not need to use a proprietary library, framework, or tool. 
You do not even have to dig into graph grammar theory.

\keywords{Event Sourcing  \and Triple Graph Grammars \and Bidirectional Model Transformations.}
\end{abstract}

\section{Introduction}

Whenever you have two tools that each deploy their own meta model 
but interchange related or 
overlapping data you face the problem of model synchronisation: 
whenever some common data is 
modified in one tool, you want to update the corresponding data in the other tool. 
Note, when both models use the same meta model, the problem of model synchronisation 
becomes closely related to model versioning and to the merging of concurrent model 
changes and to collaborative editing. 

In the 
area of bidirectional (BX) transformations there are various approaches attacking the
problem of model synchronisation 
cf. \cite{czarnecki2009bidirectional,anjorin2017benchmarx}.
Among the various approaches, we consider Triple Graph Grammars (TGGs) \cite{schurr1994specification} 
to be the most mature and most practical solution with a lot of tool support 
\cite{hildebrandt2013survey}. Recent development in TGG tools provide support for 
incremental model synchronisation \cite{leblebici2014comparison}, i.e. the effort for 
model synchronisation is proportional to the model change performed. In  
\cite{fritsche2020avoiding} Fritsche et al present recent advances in achieving 
incremental model synchronisation. 

While TGGs have a lot of mature tool support and a very sound theory, it is quite complex to 
implement a TGG tool and to apply a TGG approach within your own application. You will probably 
fail to implement your own approach and you will need to use some existing tool that requires you 
to adopt a lot of tool specific prerequisites (e.g. the Eclipse Modeling Framework 
\cite{steinberg2008emf}) and to learn a lot about (triple) graph grammar theory in order to 
write down appropriate TGG rules. 

This paper proposes Commutative Event Sourcing (CES) as an alternative approach to model synchronisation. 
As we will discuss, CES is a restricted form of a TGG where the order of rule application 
is commutative. This facilitates the implementation of CES tremendously, such that you may adopt 
our approach without the need of using a proprietary tool and without graph grammar theory. 
You just follow some design patterns that we propose and implement your own incremental model 
synchronisation manually.

\section{Triple Graph Grammars (TGGs)} \label{sec.TGGs}

\begin{figure}[hp] \centering
	\includegraphics[width=\linewidth]{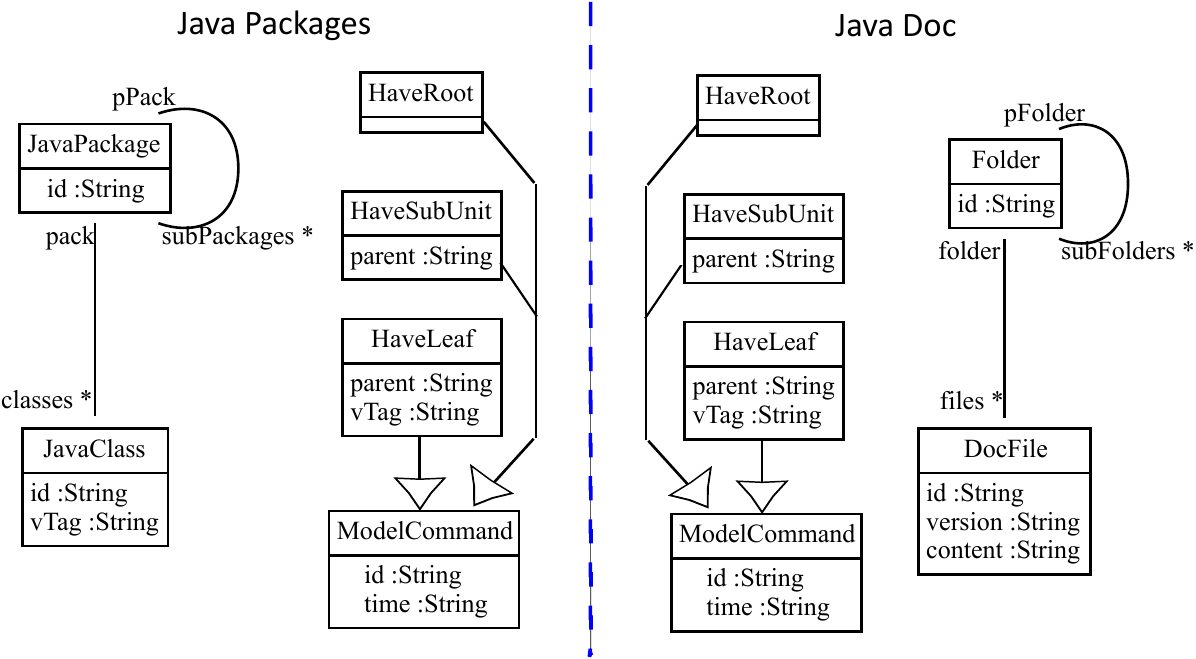}
 \caption{Classes for Java and JavaDoc structures plus edit commands}
 \label{fig:exampleClassModel}
\end{figure}

This section revisits the work of Fritsche et al \cite{fritsche2020avoiding}. 
The running example of \cite{fritsche2020avoiding} and of this paper 
is the synchronisation of Java package and Java class models with 
JavaDoc folders and files. Figure~\ref{fig:exampleClassModel} 
shows the class diagrams for the two models as 
used in our implementation of this example. 

The left of Figure~\ref{fig:exampleClassModel} shows the class model for the Java packages tool. 
Basically there is the class \texttt{JavaPackage} that may have multiple \texttt{subPackages}. 
In addition, a \texttt{JavaPackage} may have many \texttt{classes} of type \texttt{JavaClass}. 
Our class model extends the example from \cite{fritsche2020avoiding} with \texttt{id} 
attributes and with a \texttt{vTag} attribute. The latter will be used to discuss some 
editing or merge conflicts that are not handled by \cite{fritsche2020avoiding}. 
The \texttt{id} attributes are used for referencing across tools. 
The \texttt{id} attributes are
the first means that we introduce in order to facilitate the model synchronisation task. 

The left of Figure~\ref{fig:exampleClassModel} also shows the \texttt{ModelCommand} classes
\texttt{HaveRoot}, \texttt{HaveSub\-Unit}, and \texttt{HaveLeaf}. 
These classes are part of our 
CES approach and will be discussed in Section~\ref{sec.exampleImplentation}.
Our approach uses the Command pattern from \cite{gamma1995design}. Thus, we have command classes
that provides methods for command execution and \textit{command objects} that protocol each command 
execution and its actual parameters. When we serialize these command objects and exchange them with 
other applications we frequently call them \textit{events}. Thus, if we receive an event and 
deserialize it, it becomes a command object that then may be executed. In the following we will 
use command and event in a mixed way. 
The right of Figure~\ref{fig:exampleClassModel} shows the classes \texttt{Folder} and 
\texttt{DocFile} with the associations \texttt{subFolders} and \texttt{files} 
that model the JavaDoc structures. Again we have the same \texttt{ModelCommand} classes as 
for Java packages.

\begin{figure}[hp] \centering
	\includegraphics[width=\linewidth]{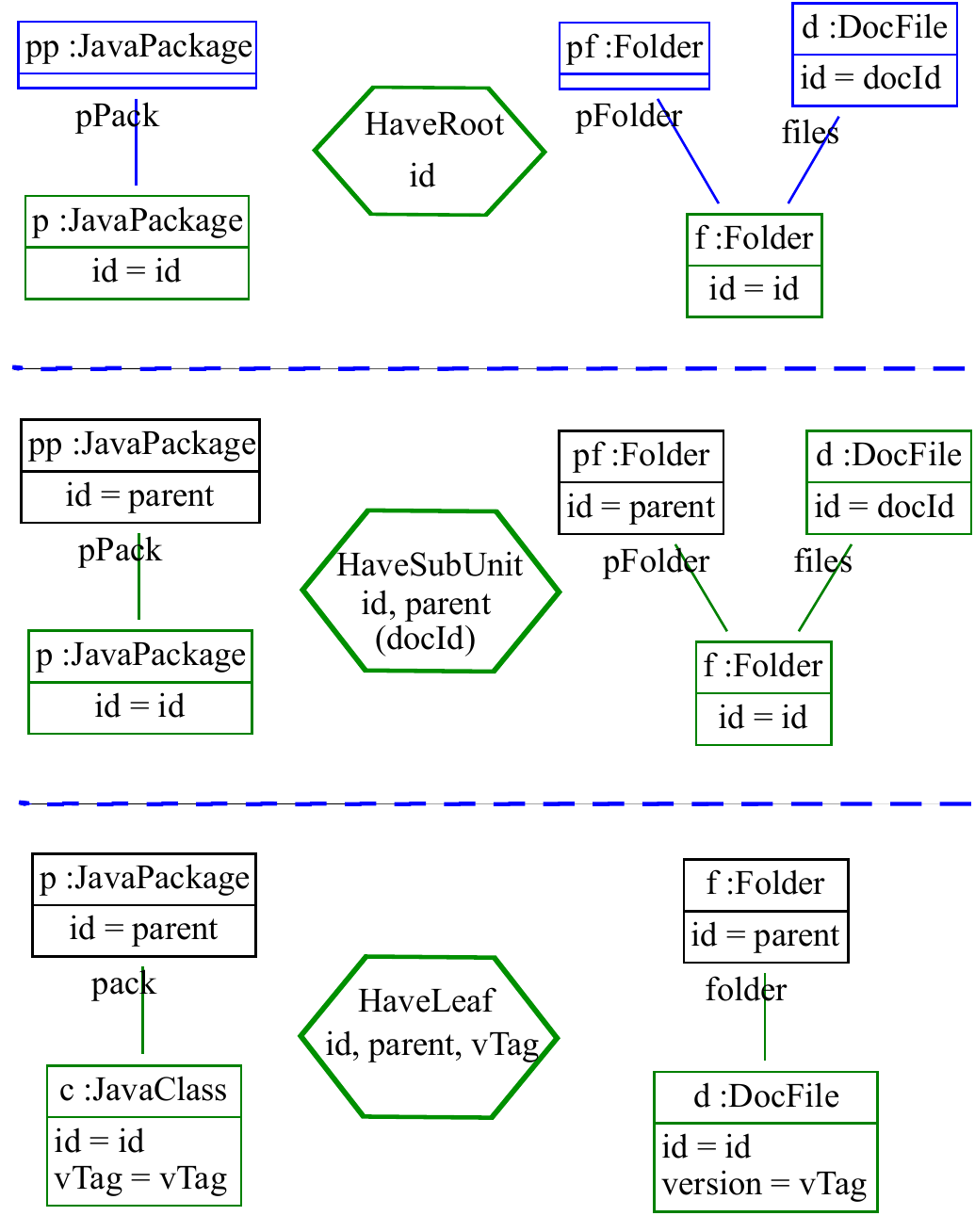}
 \caption{Triple Graph Grammar (like) rules for Java and JavaDoc structures}
 \label{fig:examplePatterns}
\end{figure}

Figure~\ref{fig:examplePatterns} shows a slight modification of the Triple Graph Grammar rules 
used in \cite{fritsche2020avoiding} to solve the model synchronisation problem for our Java 
to JavaDoc example. There are a \texttt{HaveRoot}, a \texttt{HaveSubUnit}, 
and a \texttt{HaveLeaf} rule. Each rule shows its name and 
its parameters in a hexagon in the 
middle of the rule. Each rule has a left and a right subrule. 
Each subrule specifies three possible operations: \texttt{run}, \texttt{remove}, 
and \texttt{parse}. We will walk through these operation types one by one. 

The \texttt{run} operation for a subrule tries to create the specified situation. 
Thus, all green parts of the subrule are going to be created, all black parts 
are required to already exist, and all blue parts must not exist (or may need 
to be removed by the \texttt{run} operation). 
Thus, the left subrule of the 
\texttt{HaveRoot} rule describes that on execution a \texttt{JavaPackage} object \texttt{p}
shall be created in the Java model. The \texttt{id} attribute of \texttt{p} is copied from 
the \texttt{id} parameter of the rule. The blue parts of the \texttt{HaveRoot} rule
require that there must be no \texttt{JavaPackage} \texttt{pp} attached to \texttt{p} 
via a \texttt{pPack} link. Thus we shall reset this link.  
Our manual implementation of this subrule is shown in Listing~\ref{JavaPackage.HaveRoot.run} 
See \cite{fulibServiceGeneratorGithub} for a complete reference.

\begin{lstlisting}[language=Java, numbers=left, captionpos=b, 
label={JavaPackage.HaveRoot.run}, escapeinside={\%}{\%},
caption={Manual implementation of HaveRoot rule for Java packages}
]
package JavaPackages;
public class HaveRoot extends ModelCommand {
   @Override
   public Object run(JavaPackagesEditor editor) {
      JavaPackage p = (JavaPackage) editor
            .getOrCreate(JavaPackage.class, this.getId());
      p.setPPack(null);
      return p;
   }
   ...
\end{lstlisting}

The \texttt{run} method of our \texttt{HaveRoot} command uses an \texttt{editor} to
\texttt{getOrCreate} the desired \texttt{JavaPackage} object. 
Our editor maintains a hash table 
storing \texttt{id} - \texttt{object} pairs, cf. Section~\ref{sec.getOrCreate}. 
This hash table is e.g. used in the \texttt{HaveSubUnit}
command to look up the required \texttt{pp} \texttt{JavaPackage}, 
cf. method \texttt{getObjectFrame} in 
Line~8 of Listing~\ref{JavaPackage.HaveSubUnit.run}
and Section~\ref{sec.getOrCreate}. 
Note, in the left subrule of the 
\texttt{HaveSubUnit} rule of Figure~\ref{fig:examplePatterns} the upper \texttt{JavaPackage} 
\texttt{pp} is shown in black color. This means the execution of the subrule 
requires that \texttt{pp} exists as context for the creation of the sub package 
\texttt{p}. The green \texttt{pPack} link requires that \texttt{p} needs to be connected 
to \texttt{pp} via an \texttt{pPack} link (which simultaneously creates the 
\texttt{subPackages} link in the reverse direction).

\begin{lstlisting}[language=Java, numbers=left, captionpos=b, 
label={JavaPackage.HaveSubUnit.run}, escapeinside={\%}{\%},
caption={Manual implementation of HaveSubUnit rule for Java packages}
]
package JavaPackages;
public class HaveSubUnit extends ModelCommand {
   @Override
   public Object run(JavaPackagesEditor editor) {
      JavaPackage p = (JavaPackage) editor
            .getOrCreate(JavaPackage.class, this.getId());
      JavaPackage pp = (JavaPackage) editor
            .getObjectFrame(JavaPackage.class, this.parent);
      p.setPPack(pp);
      return p;
   }
   ...
\end{lstlisting}

For completeness, Listing~\ref{JavaPackage.HaveLeaf.run} shows the \texttt{HaveLeaf}
command.

\begin{lstlisting}[language=Java, numbers=left, captionpos=b, 
label={JavaPackage.HaveLeaf.run}, escapeinside={\%}{\%},
caption={Manual implementation of HaveLeaf rule for Java packages}
]
package JavaPackages;
public class HaveLeaf extends ModelCommand {
   @Override
   public Object run(JavaPackagesEditor editor)
   {
      JavaClass c = (JavaClass) editor
            .getOrCreate(JavaClass.class, this.getId());
      JavaPackage p = (JavaPackage) editor
            .getObjectFrame(JavaPackage.class, this.parent);
      c.setPack(p);
      c.setVTag(this.vTag);
      return c;
   }
   ...
\end{lstlisting}

\begin{figure}[ht] \centering
	\includegraphics[width=\linewidth]{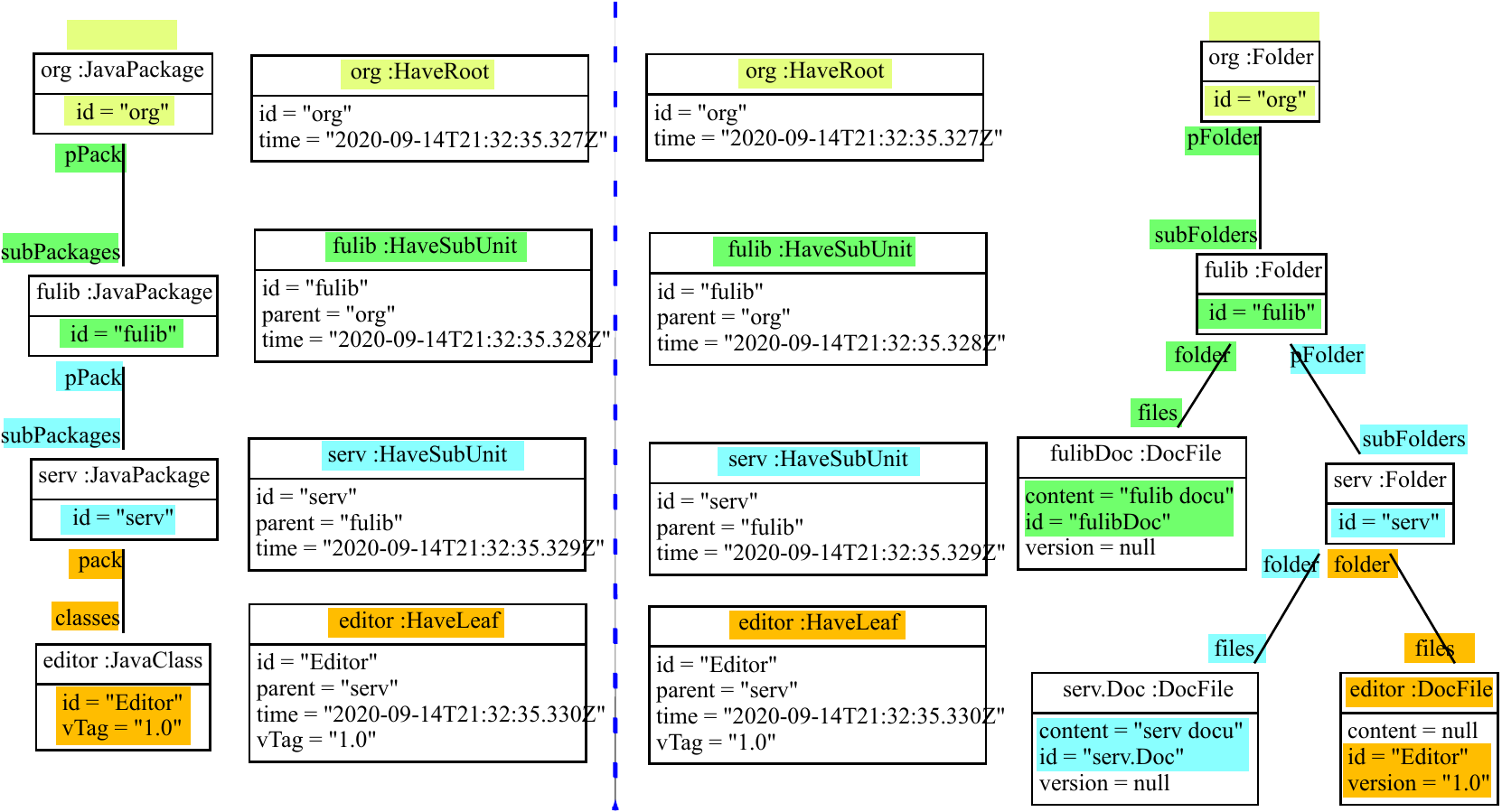}
 \caption{Objects for Java and JavaDoc structures plus commands 
          (colors cf. Section~\ref{sec.exampleImplentation})}
 \label{fig:exampleObjects}
\end{figure}

In Listing~\ref{Test.start} we create a number of command objects and initialize 
their parameters, appropriately. Then we ask an appropriate editor to execute 
the commands. The editor adds the commands to its command store and then calls 
their \texttt{run} method, cf. Section~\ref{sec.exampleImplentation}. 
This results in the object structure shown on the 
left of Figure~\ref{fig:exampleObjects}.

\newpage

\begin{lstlisting}[language=Java, numbers=left, captionpos=b, 
label={Test.start}, escapeinside={\%}{\%},
caption={Invoking triple rules or commands}
]
package javaPackagesToJavaDoc;
...
public class TestPackageToDoc {
   private void startSituation(JavaPackagesEditor editor) {
      ModelCommand cmd = new HaveRoot().setId("org");
      editor.execute(cmd);
      cmd = new HaveSubUnit().setParent("org").setId("fulib");
      editor.execute(cmd);
      cmd = new HaveSubUnit().setParent("fulib").setId("serv");
      editor.execute(cmd);
      cmd = new HaveLeaf().setParent("serv")
            .setVTag("1.0").setId("Editor");
      editor.execute(cmd);
   }
   ...
\end{lstlisting}

Listing~\ref{JavaDoc.HaveRoot.run} shows our manual implementation of the 
\texttt{HaveRoot} command for \texttt{JavaDoc} structures. This implements 
the right subrule of the \texttt{HaveRoot} triple rule of Figure~\ref{fig:examplePatterns}.
Lines~5 to 7 of Listing~\ref{JavaDoc.HaveRoot.run} are quite similar to the 
corresponding \texttt{JavaPackages} command. They just create a \texttt{Folder} instead of 
a \texttt{JavaPackage}. Lines~8 to 13 of Listing~\ref{JavaDoc.HaveRoot.run} remove 
a potentially existing \texttt{DocFile} \texttt{d}. Such an object \texttt{d} might have 
been created by previous command executions. The blue parts of the right subrule of our 
\texttt{HaveRoot} triple rule require that after rule execution 
such a \texttt{DocFile} must not (no longer) exist. 
Listing~\ref{JavaDoc.HaveSubUnit.run} and Listing~\ref{JavaDoc.HaveLeaf.run} show 
the manual implementation of the other two \texttt{JavaDoc} subrules. 

\begin{lstlisting}[language=Java, numbers=left, captionpos=b, 
label={JavaDoc.HaveRoot.run}, escapeinside={\%}{\%},
caption={Manual implementation of HaveRoot rule for JavaDoc}
]
package JavaDoc;
public class HaveRoot extends ModelCommand {
   @Override
   public Object run(JavaDocEditor editor) {
      Folder f = (Folder) editor
            .getOrCreate(Folder.class, this.getId());
      f.setPFolder(null);
      String docId = this.getId() + ".Doc";
      DocFile d = f.getFromFiles(docId);
      if (d != null) {
         editor.removeModelObject(d.getId());
         f.withoutFiles(d);
      }
      return f;
   }
   ...
\end{lstlisting}

\begin{lstlisting}[language=Java, numbers=left, captionpos=b, 
label={JavaDoc.HaveSubUnit.run}, escapeinside={\%}{\%},
caption={Manual implementation of HaveSubUnit rule for JavaDoc}
]
package JavaDoc;
public class HaveSubUnit extends ModelCommand {
   @Override
   public Object run(JavaDocEditor editor) {
      Folder f = (Folder) editor
            .getOrCreate(Folder.class, this.getId());
      Folder pf = (Folder) editor.
            getObjectFrame(Folder.class, parent);
      f.setPFolder(pf);
      String docId = this.getId() + ".Doc";
      DocFile d = (DocFile) editor
            .getOrCreate(DocFile.class, docId);
      d.setContent(this.getId() + " docu");
      f.withFiles(d);
      return f;
   }
   ...
\end{lstlisting}

\begin{lstlisting}[language=Java, numbers=left, captionpos=b, 
label={JavaDoc.HaveLeaf.run}, escapeinside={\%}{\%},
caption={Manual implementation of HaveLeaf rule for JavaDoc}
]
package JavaDoc;
public class HaveLeaf extends ModelCommand {
   @Override
   public Object run(JavaDocEditor editor) {
      DocFile d = (DocFile) editor
            .getOrCreate(DocFile.class, this.getId());
      Folder f = (Folder) editor
            .getObjectFrame(Folder.class, parent);
      d.setFolder(f);
      d.setVersion(vTag);
      return d;
   }
   ...
\end{lstlisting}

We might invoke the \texttt{JavaDoc} commands as we have done this for 
the \texttt{Java\-Packages} in Listing~\ref{Test.start}. Alternatively, 
in Line~12 of Listing~\ref{Test.firstSync} we lookup the list of commands 
that our \texttt{javaPackagesEditor} has collected while building the 
start situation. Then Line~13 uses a simple \texttt{Yaml} encoder to 
serialize these commands into a string in Yaml format. You may use any 
JSON based serialization, either. The serialization task is pretty simple, 
as our commands use string based parameters only and have no references to 
other objects. Finally, Line~14 calls method \texttt{loadYaml} on the 
\texttt{javaDocEditor}. Method \texttt{loadYaml} turns the passed string 
into \texttt{JavaDoc} command objects and executes these. The result is shown 
on the right of Figure~\ref{fig:exampleObjects}.

\begin{lstlisting}[language=Java, numbers=left, captionpos=b, 
label={Test.firstSync}, escapeinside={\%}{\%},
caption={Invoking triple rules or commands}
]
package javaPackagesToJavaDoc;
...
public class TestPackageToDoc {
   @Test
   public void testFirstForwardExample()
   {
      JavaPackagesEditor javaPackagesEditor = 
            new JavaPackagesEditor();
      startSituation(javaPackagesEditor);
      JavaDocEditor javaDocEditor = new JavaDocEditor();
      Collection commands = javaPackagesEditor
            .getActiveCommands().values();
      String yaml = Yaml.encode(commands);
      javaDocEditor.loadYaml(yaml);
      ...
   }
   ...
\end{lstlisting}

In a Triple Graph Grammar (TGG) tool based approach, you would just provide the 
triple rules shown in Figure~\ref{fig:examplePatterns}. Then the corresponding 
commands would be derived using either a code generator or a TGG interpreter. 
See \cite{fulibServiceGeneratorGithub} for an example of such an interpreter. 
Using a TGG tool, you get not only the forward execution of rules but also 
\texttt{remove} and \texttt{parse} functionality. 
Removing the effects of a TGG subrule basically requires 
to remove all model parts that have been created for green rule elements on 
the forward execution. 
In a manual implementation, you have to implement the \texttt{remove} step 
yourself and it must be consistent to the \texttt{run} operation. 
Listings~\ref{JavaPackage.HaveRoot.remove}, \ref{JavaPackage.HaveSubUnit.remove},
and \ref{JavaPackage.HaveLeaf.remove} show our manual implementation of the 
\texttt{remove} operations for the \texttt{JavaPackages} commands. 
For the \texttt{JavaDoc} commands see~\cite{fulibServiceGeneratorGithub}.

\begin{lstlisting}[language=Java, numbers=left, captionpos=b, 
label={JavaPackage.HaveRoot.remove}, escapeinside={\%}{\%},
caption={Manual implementation of HaveRoot.remove() for Java packages}
]
package JavaPackages;
public class HaveRoot extends ModelCommand {
   ...
   @Override
   public void remove(JavaPackagesEditor editor)
   {
      editor.removeModelObject(this.getId());
   }
   ...
\end{lstlisting}

\begin{lstlisting}[language=Java, numbers=left, captionpos=b, 
label={JavaPackage.HaveSubUnit.remove}, escapeinside={\%}{\%},
caption={Manual implementation of HaveSubUnit.remove() for Java packages}
]
package JavaPackages;
public class HaveSubUnit extends ModelCommand {
   ...
   @Override
   public void remove(JavaPackagesEditor editor)
   {
      JavaPackage p = (JavaPackage) editor
            .removeModelObject(this.getId());
      p.setPPack(null);
   }
   ...
\end{lstlisting}

\begin{lstlisting}[language=Java, numbers=left, captionpos=b, 
label={JavaPackage.HaveLeaf.remove}, escapeinside={\%}{\%},
caption={Manual implementation of HaveLeaf.remove() for Java packages}
]
package JavaPackages;
public class HaveLeaf extends ModelCommand {
   ...
   @Override
   public void remove(JavaPackagesEditor editor)
   {
      JavaClass c = (JavaClass) editor
            .removeModelObject(this.getId());
      c.setPack(null);
   }
   ...
\end{lstlisting}

Usually, TGG rule applications depend on each other. For example if 
we call \texttt{remove} on the \texttt{JavaPackages} \texttt{HaveRoot} command of our example, 
this would remove the \texttt{org} \texttt{JavaPackage} from our model, cf. 
Figure~\ref{fig:exampleObjects}. This would leave the \texttt{fulib} 
\texttt{JavaPackage} without a parent. Thus the \texttt{HaveSubUnit} TGG rule 
does no longer match for the \texttt{fulib} object. This is important as 
on model synchronisation the right sub rule of \texttt{HaveSubUnit} creates the 
\texttt{fulibDoc} \texttt{DocFile} which is no longer valid. To repair this, 
a standard TGG approach has to \texttt{remove} dependant rule applications whenever their 
context becomes invalid. Thus on \texttt{remove} of the \texttt{HaveRoot} command, 
a standard repair step would also \texttt{remove} the \texttt{HaveSubUnit} command for the 
\texttt{fulib} \texttt{JavaPackage} and in turn \texttt{remove} the commands for the 
\texttt{serv} and for the \texttt{editor} objects. Via model synchronisation all 
\texttt{JavaDoc} objects will be removed, either. To keep the lower model parts, 
one would apply a \texttt{HaveRoot} command on \texttt{fulib} and rerun the 
\texttt{HaveSubUnit} and \texttt{HaveLeaf} commands on \texttt{serv} 
and \texttt{editor}. 
Formally, the whole model would be deleted and reconstructed. 
In \cite{fritsche2020avoiding} 
this is called a cascading delete and \cite{fritsche2020avoiding} proposes 
sophisticated theory and means to avoid this cascading delete via so called 
short-cut-repair rules.
In our manual implementation it would suffice to \texttt{run} a \texttt{HaveRoot}
command on \texttt{fulib} in order to repair the situation. This is achieved as 
Line~7 of Listing~\ref{JavaPackage.HaveRoot.run} and Lines~7 to 13 of 
Listing~\ref{JavaDoc.HaveRoot.run} carefully remove all model parts that 
correspond to the blue parts of the \texttt{HaveRoot} TGG rule, cf. 
Figure~\ref{fig:examplePatterns}, i.e. all model parts that might stem from 
a previous application of a \texttt{HaveSubUnit} rule. In a manual implementation 
you have to spot the overlap of the \texttt{HaveRoot} and the \texttt{HaveSubUnit}
rules yourself and you have to design theses rules and their manual implementation 
very carefully in order to circumvent cascading deletes. 
Commutative Event Sourcing will help you to achieve this as discussed 
in Sections~\ref{sec.CESTheory} and \ref{sec.exampleImplentation}.
\cite{fritsche2020avoiding} does this for you automatically which is 
a really great job. 

Whenever you edit a model directly without using the TGG rules or the 
corresponding commands and you want to do a new model synchronisation, 
you need to parse the changed model in order to 
identify which TGG rule applications are now valid. Again TGG tools do this
parsing for you. Basically all green and black parts of a TGG subrule must 
be matched and the blue parts must not be there. In general TGG parsing has to deal with 
rule dependencies, too. Usually, TGG parsing needs to find all applications of 
so-called "root" rules (that do not depend on other rules), first. 
Then TGG parsing, inspects the surroundings of "root" rule applications and 
tries to find applications of rules that use the "root" rules in their context. 
In turn, you apply rules where the context has become available.
In our example this results in some kind of top-down parsing starting with 
the \texttt{org} \texttt{JavaPackage} and descending to \texttt{subPackages}
and \texttt{classes}, recursively. In general, TGG parsing may be even more 
complicated, cf. \cite{schurr1994specification}.
Again Commutative Event Sourcing 
allows us to facilitate parsing considerably as we will discuss 
in Section~\ref{sec.parsing}. 

Thus in our manual implementation we ignore the order of rule applications for now. 
For a single rule, parsing is relatively simple. 
Listing~\ref{JavaPackage.HaveRoot.parse} shows the manual implementation of the 
\texttt{parse} method for our \texttt{HaveRoot} command for \texttt{JavaPackages}. 
Our editor calls the \texttt{parse} methods of our commands when appropriate, 
cf. Section~\ref{sec.parsing}. On such a call, 
the editor passes an object that may have been modified 
and needs parsing as parameter. 
Thus Line~6 of the \texttt{parse} method of 
Listing~\ref{JavaPackage.HaveRoot.parse} first ensures that the current object
is a \texttt{JavaPackage}. Similarly, Line~10 ensures that there is no \texttt{pPack}. 
If the current package has no sub packages and contains no 
\texttt{JavaClass}es, we consider it garbage. Therefore, Line~15 to 17 return 
a \texttt{RemoveCommand} with the corresponding object \texttt{id}. Without such 
a garbage collection mechanism, the users would have to invoke \texttt{RemoveCommand}s
manually in order to get rid of model objects. Finally Line~20 to 22 create a 
\texttt{HaveRoot} command and retrieve its \texttt{id} parameter from the model 
and return it. 

\begin{lstlisting}[language=Java, numbers=left, captionpos=b, 
label={JavaPackage.HaveRoot.parse}, escapeinside={\%}{\%},
caption={Manual implementation of the parse step for HaveRoot for Java packages}
]
package JavaPackages;
public class HaveRoot extends ModelCommand {
   ...
   @Override
   public ModelCommand parse(Object currentObject) {
      if (! (currentObject instanceof JavaPackage)) {
         return null;
      }
      JavaPackage currentPackage = (JavaPackage) currentObject;
      if (currentPackage.getPPack() != null) {
         return null;
      }
      if (currentPackage.getClasses().isEmpty() 
            && currentPackage.getSubPackages().isEmpty()) {
         ModelCommand modelCommand = new RemoveCommand()
               .setId(currentPackage.getId());
         return modelCommand;
      }
      // yes its me
      ModelCommand modelCommand = new HaveRoot()
            .setId(currentPackage.getId());
      return modelCommand;
   }
   ...
\end{lstlisting}

Listing~\ref{JavaPackage.HaveSubUnit.parse} shows the \texttt{parse} method 
of the \texttt{HaveSubUnit} command for \texttt{JavaPackages}. Note, that this 
method is slightly simpler as the garbage collection is done by the
\texttt{HaveRoot} command. You find the \texttt{parse} method of the \texttt{HaveLeaf}
command for \texttt{JavaPackages} in \cite{fulibServiceGeneratorGithub}. We leave 
the implementation of the parse methods for the \texttt{JavaDoc} commands as an 
exercise for the interested reader.

\begin{lstlisting}[language=Java, numbers=left, captionpos=b, 
label={JavaPackage.HaveSubUnit.parse}, escapeinside={\%}{\%},
caption={Manual implementation of the parse step for HaveSubUnit for Java packages}
]
package JavaPackages;
public class HaveSubUnit extends ModelCommand {
   ...
   @Override
   public ModelCommand parse(Object currentObject) {
      if ( ! (currentObject instanceof JavaPackage)) {
         return null;
      }
      JavaPackage currentPackage = (JavaPackage) currentObject;
      if (currentPackage.getPPack() == null) {
         return null;
      }
      ModelCommand modelCommand = new HaveSubUnit()
            .setParent(currentPackage.getPPack().getId())
            .setId(currentPackage.getId());
      return modelCommand;
   }
   ...
\end{lstlisting}

\section{Commutative Event Sourcing (CES) Theory} \label{sec.CESTheory}

Event Sourcing has been proposed by \cite{evans2004domain} and 
\cite{vernon2013implementing} as a 
means for communication between multiple domains or (micro) services. 
Event Sourcing may also be used as mechanism for model persistence. 
Basically, a program or service 
logs relevant operations as events  
and these events are then transferred to other 
programs or services that 
react with appropriate operations 
on their site. In order to use this 
idea for model synchronisation, 
we just log all editing operations / commands on one model and then send these 
editing events to some other 
model and perform similar changes there. This is clearly related to 
the Triple Graph Grammar approach discussed in Section~\ref{sec.TGGs}. 
However, Event Sourcing has some additional requirements that ultimately
lead us to Commutative Event Sourcing. 

\begin{figure}[ht] \centering
	\includegraphics[width=\linewidth]{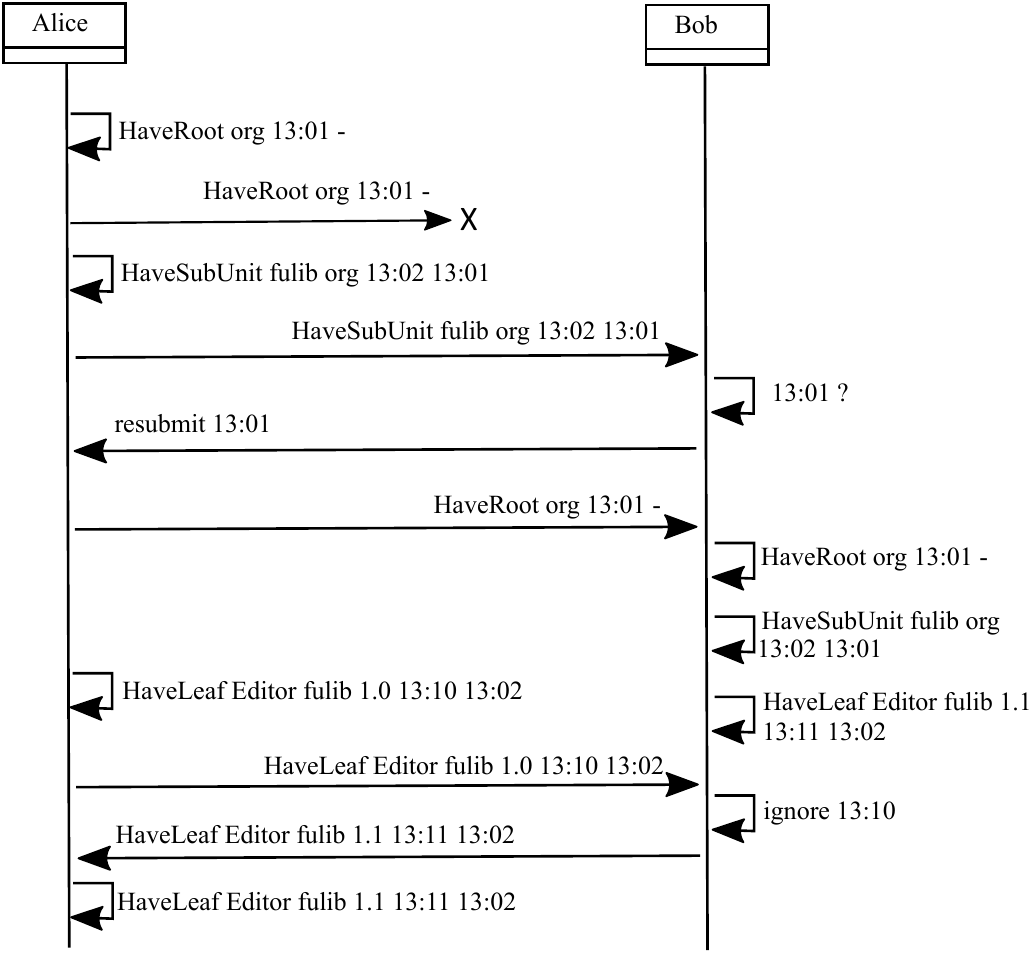}
 \caption{Collaborative Editing}
 \label{fig:CollaborativeEditing}
\end{figure}

To connect multiple applications, Event Sourcing is frequently based on 
some message broker mechanism. Depending on the quality of service that 
your message broker provides, messages may be lost or received in wrong order
and messages may be received multiple times. For example, there may be a 
(\texttt{HaveRoot org}) event and a (\texttt{HaveSubUnit fulib org}) event and 
for some reasons you receive only the latter, cf. Figure~\ref{fig:CollaborativeEditing}. 
In \cite{schneider2004coobra} 
we deal with this problem by adding \texttt{time stamps} 
to all events / commands and by extending
each event with the \textit{time stamp} of its predecessor. 
Thus if you receive event 
(\texttt{HaveSubUnit fulib org 13:02 13:01}) which has been raised at 13:02 
and which has a predecessor command that has been raised at 13:01 and you have not
yet received the 13:01 command, then 
you postpone the execution of the \texttt{13:02} event and ask for re-submission of
the \texttt{13:01} event, cf. Figure~\ref{fig:CollaborativeEditing}. 
Time stamps also allow to detect duplicated receipt of the same 
event. In addition, if there are two independent tools 
(e.g. the editors of Alice and Bob) that raise event 
(\texttt{HaveLeaf Editor fulib 1.0 13:10 13:02}) and 
(\texttt{HaveLeaf Editor fulib 1.1 13:11 13:02}) you 
have a merge conflict and the time stamps allow you 
to resolve such merge conflicts, deterministically. 

While time stamps are a good idea, in \cite{schneider2004coobra} 
postponing command execution until all predecessor commands have arrived was quite tricky: 
you have to detect the missing of a predecessor, you have to ask for resubmission
you have to wait for the predecessor to arrive, there may be a cascade of 
pre-predecessors you also have to ask and to wait for, finally you apply the commands 
in their correct order. Well, unless there is collaborative editing of multiple editors:
with collaborative editing, when you receive a command with time stamp 
13:30 (and predecessor 13:11) and you already got the 13:11 event, 
there still might be e.g. an event with time stamp 13:23 
(and also predecessor 13:11) that just did not reach you, yet. 
Thus, to execute collaborative commands in a correct timely order 
opens a new can of worms. 

To overcome the problem of rule ordering, we want our commands to be executed in 
any order, i.e. to become commutative. 
Thus, we want to get rid of TGG rule dependencies. 
One reason for dependencies between TGG 
rule applications is that rule execution requires that some context (black) parts 
must already exist in order to connect new elements to their context. For example, 
a (\texttt{HaveLeaf Editor serv}) command for \texttt{JavaPackages} needs to connect the 
new \texttt{Editor} \texttt{JavaClass} to an (existing) \texttt{serv} \texttt{JavaPackage}
via a \texttt{pack - classes} link, cf. Figure~\ref{fig:examplePatterns} and 
Listing~\ref{JavaPackage.HaveLeaf.run}.  
In our approach we are able to
execute the \texttt{HaveLeaf Editor serv} command 
before the \texttt{serv} \texttt{JavaPackage} is created. 
We achieve this by using \texttt{id}s for 
model objects and by having a hash table of all model objects and by using  
\texttt{getOrCreate} and \texttt{getObjectFrame} operations
that do a hash table lookup for the required object (\texttt{id}) 
and create the object if it is missing, cf. Line~9 of 
Listing~\ref{JavaPackage.HaveLeaf.run}. 
The details of the \texttt{getOrCreate} and \texttt{getObjectFrame} mechanism
are explained in Section~\ref{sec.parsing}. 
Thus, our (\texttt{HaveLeaf Editor serv}) command creates a context 
object with \texttt{id} \texttt{serv} and creates the \textit{pack} link to it.
When we execute the (\texttt{HaveSubUnit serv fulib}) command later, it uses 
\texttt{getOrCreate(serv)} to retrieve the \texttt{serv} 
\texttt{JavaPackage} that has already been created by our \texttt{HaveLeaf} 
command and to connect the \texttt{serv} \texttt{JavaPackage} to 
its \texttt{pPack} \texttt{fulib}. The \texttt{fulib} \texttt{JavaPackage}
is again retrieved (or created) via \texttt{getObjectFrame}. 

This \texttt{getOrCreate} mechanism is inspired by QVT Relations \cite{qvt}
and \cite{macedo2013implementing},
however QVT Relations allows 
to combine any attributes that may be used as keys while we restrict this 
to just \texttt{id}s, in order to facilitate a manual implementation. 
In addition, QVT Relations still deploys a hierarchy of rules which requires 
considerable implementation effort for the handling of rule dependencies.

Due to our \texttt{getOrCreate}, 
mechanism together with some additional rule restrictions that are 
discussed in Section~\ref{sec.exampleImplentation}, 
in our approach it is possible to execute 
commands in any order, i.e. our commands are commutative. 
Having commutative commands gets rid of the effort for command ordering. 
And it facilitates parsing, cf. Section~\ref{sec.parsing}. 

There is one additional problem with general event sourcing: usually your event 
store grows over time. For example if you have a 
(\texttt{HaveLeaf Editor serv 1.1}) command that creates an \texttt{Editor} 
object and assigns \texttt{1.1} to its \texttt{vTag} and than you have a 
new (\texttt{HaveLeaf Editor serv 1.2}) command that just changes the 
\texttt{vTag} to \texttt{1.2} and later on you change \texttt{vTag} again 
and again. Now your event store may contain a large number of 
\texttt{HaveLeaf} events that only differ in their \texttt{vTag} parameter. 
Actually, it would suffice to keep only the latest \texttt{HaveLeaf} 
command in order to recreate the final model or to synchronize with some 
other model. In our approach, we want a simple mechanism to identify events 
that overwrite each other in order to be able to restrict the size of 
our event store to be proportional to the size of our model. 
Therefore, our implementation uses command \texttt{id}s and in our implementation
two commands that have the same \texttt{id} parameter must overwrite each other's effects
such that it suffices to keep one event per command \texttt{id} in our store, 
cf. Section~\ref{sec.exampleImplentation} and Listing~\ref{JavaPackage.Editor.execute}. 

Next, we 
provide some theory that specifies the requirements for Commutative Event Sourcing
and compressed event histories, formally. 
Section~\ref{sec.exampleImplentation} then shows simple patterns that achieve 
a sound implementation of Commutative Event Sourcing. 
A previous version of our theory has been published in \cite{copei2019mx}. 
However, this paper restructures our theory in large parts and it even 
enriches our formal requirements in order to further facilitate the implementation 
of a Commutative Event Sourcing application. 

First let us set up some basic notations: like \cite{stevens2010bidirectional} 
we use capital letters such as $M$, $N$ for metamodels i.e. for sets of models 
(that adhere to a common class diagram). $\emptyset$ denotes an empty model. 
We denote events with $e$ and the set of all possible events with $E$. 
An event $e = (t, i, x_1, ..., x_n)$ has an event type $t \in T$, an 
event identifier $i \in \mathbb{CHAR^*}$ (i.e. some string),
and a number of parameters 
$x_i \in \mathbb{R} \, \cup \, \mathbb{CHAR^*}$, 
i.e. parameters are arbitrary numbers or strings. 
We will also use series of events
$\overline{e} = (e_1, ..., e_n)$ and denote by $\overline{E}$ the set of all 
possible event series with events from $E$. Similarly we use sets of events 
$\tilde{e} = \{e_1, ..., e_n\} \in \mathcal{P}(E)$.  

Events may be applied to (or synchronized with) models via the function  
$\textit{apply}(e, m)$, which generates a possibly new model $m'$. 
Furthermore, we define the application of an event series 
$\overline{e} = (e_1, \ldots, e_n)$ to a model $m$ as \newline
$\textit{apply}(\overline{e}, m) = \textit{apply}(e_n, \ldots, 
\textit{apply}(e_2, \textit{apply}(e_1, m))\ldots)$. \newline
Similarly, the application of a set of events $\tilde{e}$ is 
defined as the applications of all its elements in some order.

Now, we want a simple mechanism that allows us to restrict the size of 
our command histories. Therefore, Definition~\ref{prop.overwriting}
requires that events with the same \texttt{id} overwrite each others 
effects, i.e. one renders the other ineffective. 
This allows us to maintain our commands within a hash table and if we run a 
new command with an already used \texttt{id}, Definition~\ref{prop.overwriting}
allows us to overwrite the old hash table entry in our command store with the 
new command.
 
\begin{definition} (effective events) \label{def.efficient}
An event $e_p$ is called ineffective within an event series $\overline{e}$ at position $p$
iff $\textit{apply}(\overline{e} \setminus e_p, \emptyset) = \textit{apply}(\overline{e}, \emptyset)$.
\newline
We call $e_p$ effective, otherwise. \newline
We write  $\tilde{f} = \textit{effective}(\tilde{e})$ to denote the
series $\tilde{f}$ that is derived from $\tilde{e}$ by removing all ineffective 
events. 

\end{definition}

\begin{definition} (overwriting) \label{prop.overwriting}
We call events
$e_1 = (t_1, i_1, ...)$ and $e_2 = (t_2, i_2, ...)$ with identifier $i_1 = i_2$
\textit{overwriting} within an event series $\overline{e}$, if
the earlier event is ineffective in $\overline{e}$.  
\end{definition}

Next we want to get rid of command dependencies, i.e. we want to be able to execute
a command as soon as we receive it without waiting for other commands to establish 
a required context. Similarly, we would like to store our (unordered) command hash table persistently
and to reload and rerun all commands on tool start up without bothering with command order. 

\begin{definition} (commutativity) \label{prop.commutativity}
We call events \textit{commutative}, if
for all models $m \in M$ and all events $e_1, e_2 \in E$  
holds that \newline
$\textit{apply}(e_2, \textit{apply}(e_1, m)) = \textit{apply}(e_1, \textit{apply}(e_2, m))$
\end{definition}

Definition~\ref{prop.commutativity} just states that command execution must be commutative. 
Section~\ref{sec.exampleImplentation} will show how to achieve this in your implementation. 

\begin{definition} (active event sets $\tilde{e}$) \label{prop.active-events} \newline
For any event series $\overline{e}$ where for all $e_1, e_2$ in  $\textit{effective}(\overline{e})$ holds \newline
(1) $e_1$ and $e_2$ have distinct identifiers, (i.e. all events in 
    $\overline{e}$ are overwriting) and \newline
(2) $e_1$ and $e_2$ are commutative \newline
we define the set of active events 
$\tilde{e}$ = \{ $e$ in $\textit{effective}(\overline{e})$ \}. \newline
We write $\tilde{e} = \textit{activeSet}(\overline{e})$ to denote the set 
of active events $\tilde{e}$ derived from $\textit{effective}(\overline{e})$.

\end{definition}

We are now ready to define $M_\texttt{CES}$ 
the set of all models that may be created via Commutative Event Sourcing:

\begin{definition} ($M_\textit{CES}$) \label{prop.parse-ability}
A model $m = \textit{apply}(\overline{e}, \emptyset)$ 
is supporting Commutative Event Sourcing if and only if  \newline
for all event series $\overline{e}_1$ and $\overline{e}_2$ with \newline
(1) $m = \textit{apply}(\overline{e}_1, \emptyset)$ and 
    $m = \textit{apply}(\overline{e}_2, \emptyset)$ and 
    \newline
(2) all events with the same identifier in $\overline{e}_1$ and $\overline{e}_2$ 
    are overwriting and \newline     
(3) all events in $\textit{effective}(\overline{e}_1)$ and 
    $\textit{effective}(\overline{e}_2)$ are commutative. \newline  
holds there exists a unique set of events $\tilde{e}$ with $\tilde{e} = \textit{activeSet}(\overline{e}_1)$ and  $\tilde{e} = \textit{activeSet}(\overline{e}_2)$.
\newline
In addition we require a  
function $\textit{parse}: M \to \mathcal{P}(E)$ such that
$\textit{parse}(m) = \tilde{e}$. \newline
We say $m \in M_\textit{CES}$.
\end{definition}

This means, for any $m \in M_\textit{CES}$ 
there is a uniquely defined event set $\tilde{e}$
such that $m = \textit{apply}(\tilde{e}, \emptyset)$ and it is possible to derive 
$\tilde{e}$ from $m$ via parsing. Thus, 
we require a bidirectional mapping between the command store and its model and 
it must be possible to reconstruct the command store from the model through parsing. 

Now we are able to define whether two models are "equivalent" or synchronized. 
Two models are synchronized if their active event sets are equal. We may also opt for 
partial synchronisation e.g. we may restrict ourselves to certain event types. Then 
two models are partially synchronized if their active event sets restricted to the 
desired event types are equal. This allows e.g. to have additional
\texttt{HaveContent} commands
in our \texttt{JavaDoc} tool that add the actual content to our \texttt{JavaDoc} 
\texttt{File}s. This additional 
information is not synchronized with our \texttt{JavaPackages} tool, 
cf. \cite{fulibServiceGeneratorGithub}. 

Overall, we restrict ourselves to commands that have to be implemented in a very specific way.
However, this should not restrict the kind of models that may be generated, too severe. 
We will discuss this in our conclusions.

\section{Achieving Commutative Event Sourcing}\label{sec.exampleImplentation} 

This section provides some simple design rules to achieve Commutative Event Sourcing. 
Our design rules are based on the notion of an \textit{increment}. 
An increment is a set of 
attributes and links that are edited through a single command execution. 
With respect to our \texttt{getOrCreate} mechanism, objects are part of an 
increment if their \texttt{id} attribute is a part of the increment. 
In Figure~\ref{fig:exampleObjects} we have color coded different increments 
and their commands. 

To achieve commutativity for our commands, we first look on commutativity for TGG rules. 
Basically, a TGG rule has a context (black) part that shall already exist and that is 
not modified and a main (green and blue) part that defines its increment, i.e. 
the attributes and links that will be edited on rule execution. 
In general, a TGG rule may have an arbitrary complex context (black) part. 
You may require multiple objects that shall be connected by various links 
and you may have additional attribute constraints and application conditions. 
However, such complex context parts potentially create rule dependencies: 
for example, if a rule requires some link within its context part and later this link is 
removed, the rule application is no longer valid and it must be removed, too. This 
may cause cascading deletes. To avoid such rule dependencies, 
we restrict the context (black) part of our rules to simple 
objects with an \texttt{id} attribute constraint. No links between context 
objects and no general attribute constraints. Similarly, we restrict the negative 
context (blue) part to simple objects with id constraints. Again no links and no 
general attribute conditions. 

The main (green and blue) parts of a rule result in active changes of attribute 
values and links. This may interfere with another rule, if the other rule changes 
the same attributes or links, differently. In that case rule application order 
would matter. Thus, we require that the increment edited 
by the main rule parts must not overlap with the increment of other rule applications, 
i.e. no two rule applications shall edit the same attribute or link. 

Together, simple contexts and not overlapping main parts achieve commutativity. 
The prove of this claim is current work. It would also be great to have a compile 
time check that validates a set of triple graph grammar rules for commutativity.
In addition, we need criteria and checks for overwriting TGG rules, i.e. for rules 
that edit the same increment in different ways. 
However, if you implement the commands manually, 
such TGG criteria and checks will not apply for you. 
So far, to validate the commutativity of your TGG rules, you will have to 
execute them in different order and to check whether all orders achieve 
the same result. Due to our experience, executing a reasonable sequence of events 
once in order and once in reverse order and comparing the result works very well 
in order to detect non-commutative rules. 

During the manual implementation of our commands, it is straight forward to restrict 
ourselves to simple context (black) objects: either do not check for attribute 
values or links or if you do check an attribute, consider it as a part of your 
increment, i.e. you edit it. 

To achieve non-overlapping increments, we first introduce a convention: each command 
shall have exactly one \textit{core} object that has the same \texttt{id} as the 
command itself. Based on this convention, to find non-overlapping increments we 
start with some reasonably representative object model example. Then we choose 
some object as the core object of our first command and we mark the \texttt{id}
attribute of this object with a certain color. In Figure~\ref{fig:exampleObjects}
we may e.g. start with the \texttt{Editor} object in the lower left corner and mark 
its \texttt{id} attribute with orange. Next, we look for other attributes of the same 
object that our command may initialize in the same step and mark them with the 
same color. In our example, we choose 
the \texttt{vTag} attribute of our \texttt{Editor} object. Next, we look for to-one
links that connect our core object to other objects. In our example we choose the 
\texttt{pack - classes} link attached to our \texttt{Editor} object and mark it with 
orange, too. Now we look for other \texttt{core} objects of the same type and try to 
mark a similar increment induced by this new \texttt{core} object. 
In Figure~\ref{fig:exampleObjects} there is no other \texttt{JavaClass} object, thus
we go on with a new core object of a different type, e.g. with the \texttt{serv} object
of type \texttt{JavaPackage}. 
The \texttt{serv} object has no other attributes but a \texttt{pPack - subPackages} link
of cardinality to-one. We color the \texttt{id} attribute and the link in blue. 
Now we look at the \texttt{fulib} object on the left of Figure~\ref{fig:exampleObjects}
which is also of type \texttt{JavaPackage}. 
Here we use green color to mark a similar increment. 
If we look at the \texttt{org} object there is no \texttt{pPack - subPackages} link 
attached to it. However, there is an empty field within the \texttt{org} object that 
could hold a \texttt{pPack} link. To make the \texttt{org} increment similar to the 
other \texttt{JavaPackage} increments, we mark the empty space above the \texttt{org}
object with yellow color. Now we have colored all attributes and all links of our 
\texttt{JavaPackages} example model and no attribute and no link is marked by two colors. 
Thus, we have identified non-overlapping increments 
that cover all elements of our example model. 
So far, this approach only adds to-one associations to increments. 
In general, a model may deploy many-to-many associations. To cover a many-to-many 
association we introduce dedicated rules that consist of two context (black) nodes
and a green link between theses nodes (to create the link) or a blue link between these
nodes (to delete the link). These rules shall not contain any other elements. 
As these rules just edit a single link, they are either commutative or overwriting
by construction.  

We are now ready to define our commands: for each core object we create a command object
with a certain type. For each attribute of an increment, the command object gets an 
appropriate parameter attribute. For each link of our increment the command gets a 
parameter attribute that holds the \texttt{id} of the target object. Usually, we use 
one command type for each increment type or for each core object type. If there are two 
core objects of the same type 
which handle the attributes and links of their increment in different ways, 
you might use two different command types or one command type which distinguishes 
the two cases, internally. In our example, we use command type \texttt{HaveRoot} for 
the \texttt{org} object and command type \texttt{HaveSubUnit} for the \texttt{fulib}
and the \texttt{serv} object. These two commands differ in the \texttt{parent} parameter 
and in their handling of the \texttt{pPack} link: the former deletes the 
\texttt{pPack} link and the latter creates it. 
Generally, increments (with core objects) of the same type must be similar, 
i.e. they edit similar sets of attributes and links. 
Accordingly, commands of different types that edit increments of the same type
must edit similar sets of attributes and links. Therefore, 
the \texttt{HaveRoot} command has to assign a defined value (i.e. null) to the 
\texttt{pPack} field of its core object. Actually, in a manual implementation 
we would probably use only one command type for \texttt{JavaPackage} objects, as it 
is easy to handle both cases in one implementation. However, the example stems from 
\cite{fritsche2020avoiding} and \cite{fritsche2020avoiding} uses TGG rules and with 
TGGs you need different rules for different cases. Thus, we use two command types to 
facilitate the comparison. 

Note, as all commands that edit the same increment type 
(or core objects with the same type) 
edit similar sets of attributes and links,
our commands are overwriting by construction: 
if two commands have the same 
\texttt{id}, they edit the same core object and the same increment 
and as discussed all parts of the increment 
get a well defined value. Thus, one command will overwrite all attributes and 
links that have been edited by the other command and thus it suffices to keep 
only the last command in our command store.

Let's now have a look at the \texttt{JavaDoc} example model at the right side of 
Figure~\ref{fig:exampleObjects}. Given the increments of the left model, we now 
have to identify the corresponding increments on the right model. For the orange 
increment on the bottom, this is the \texttt{Editor} \texttt{DocFile} with its 
\texttt{id} attribute, its \texttt{version} attribute and its 
\texttt{folder - files} link. In Figure~\ref{fig:exampleObjects} this is easily 
spotted by comparing object \texttt{ids} and 
attribute values and link targets and names. Usually, you have only one example 
model as a start and you construct the target model "incrementally". Thus, you 
identify which objects, attributes and links need to be added to the target model 
in order to represent the information that is provided by the source increment or 
by the corresponding command and its parameters. And you reuse the \texttt{id}s
of the source model within the target model in order to establish the desired
correspondences. 

Note, the \texttt{content} attribute
of our \texttt{Editor} object on the right is not marked with orange. Actually, 
the \texttt{content} attribute of a \texttt{DocFile} has no correspondence 
within the \texttt{JavaPackages} model and thus it will not be addressed by 
model synchronisation but we will introduce a separate \texttt{HaveContent} command
within the \texttt{JavaDoc} model later on, cf. \cite{fulibServiceGeneratorGithub}.

For the \texttt{HaveSubUnit} increments or commands the situation is somewhat more 
complicated: Within the \texttt{JavaDoc} 
model, a sub package is represented by a \texttt{Folder} object and by a special 
\texttt{DocFile} that describes the sub package. If we look e.g. at the blue
\texttt{serv} increment in Figure~\ref{fig:exampleObjects}, 
within the \texttt{JavaDoc} model this increment is 
represented by the \texttt{serv} object (of type \texttt{Folder})
and by the \texttt{serv.Doc} object (of type \texttt{DocFile}) 
and by its \texttt{content} attribute and the attached \texttt{folder - files} and 
\texttt{pFolder - subFolders} links. Thus, our increment contains two model objects, 
the core object with id \texttt{serv} and a dependent object with id \texttt{serv.Doc}.
In general, a \texttt{Folder} object may contain multiple \texttt{DocFile} objects. 
In order to identify the \texttt{DocFile} object that describes the 
\texttt{Folder} itself, our example uses the convention that the describing 
\texttt{DocFile} has an \texttt{id} that is equal to the \texttt{id} of its 
\texttt{Folder} plus a \texttt{".Doc"} suffix, cf. Figure~\ref{fig:exampleObjects} 
and Listing~\ref{JavaDoc.HaveSubUnit.run}. With this convention it is still possible 
to identify all parts of an increment by starting at some model object and collecting 
the attached parts. If you start e.g. with the \texttt{serv.Doc} object you may identify 
the corresponding \texttt{serv} object either using our naming convention or using the 
\texttt{folder - files} link. If you start with the \texttt{serv} object you use 
the naming convention to identify the \texttt{serv.Doc} object. 
As in our \texttt{JavaDoc} model a \texttt{HaveSubUnit} increment covers a 
\texttt{DocFile} object and as the \texttt{HaveRoot} increment and command 
address the same kind of increment, the \texttt{HaveRoot} commands needs to edit
(i.e. remove) a potentially attached \texttt{DocFile} object, too, cf. 
Listing~\ref{JavaDoc.HaveRoot.run}.  

Note, the \texttt{version} attribute of our \texttt{serv.Doc} object is not colored
as the corresponding \texttt{JavaPackage} and the corresponding \texttt{HaveSubUnit} 
command do not have any version information.

To summarize, you start with core objects and collect attributes, to-one links 
(and neighbors) that form an increment, i.e. that are edited together. 
For many-to-many associations you add dedicated rules. 
Core objects of the same type 
should have increments of similar structure. 
For such increments you introduce a command
with parameter attributes that correspond to attribute values and link targets. 
If there are alternative cases, you may use different command types or choose the 
desired variant from parameter values. This decision must not rely on context 
properties. 
For each increment in some source model, you identify a corresponding increment within 
the target model. Increments shall not overlap and all (relevant) parts of the 
models shall be covered. Thereby, you construct commands with simple contexts and 
non-overlapping increments, i.e. commutative commands. 

You may check the commutativity of your commands
by applying a sufficiently long series of commands once in order and once in 
reverse order and then you compare the resulting models. To our experiences, this 
reveals violations of the commutativity rule, quite reliably.  Note, due to 
our excessive use of \texttt{id}s, comparison of two models is quite easy in our case.

\section{Collaborative Editing and Merge Conflicts} \label{sec.EditorExecute}

Once you got your commands right, you still have to deal with multiple or 
concurrent edits of the same increment. Assume, you have already executed command 
(\texttt{HaveLeaf Editor serv 1.0}) and now you want to change 
the version to 1.1. Thus you run the command (\texttt{HaveLeaf Editor serv 1.1}). 
In your current editor the second command would just overwrite the first command
and everything is fine. Unfortunately, model synchronisation may fail, if your 
message broker delivers theses two commands in the wrong order. Similarly, 
you have a model synchronisation problem, if the two commands are run in two 
different editors, concurrently, and you try to synchronize afterwards. 
As the two commands are conflicting, i.e. they assign different values to the
\texttt{vTag} attribute of the \texttt{Editor} object, model synchronisation 
needs some mechanism that decides which command overwrites the other. 

As far as we are aware of, this problem is only recently addressed 
by Triple Graph Grammar tools cf. \cite{fritsche2020precedence}. Actually, \cite{fritsche2020avoiding} explicitly mentions this problem 
and puts it in future work. This paper solves this problem with the help of 
additional time stamps and with the help of command specific merge strategies. 
Listing~\ref{JavaPackage.Editor.execute} shows the 
\texttt{execute} method that we deploy in our editors. 

\begin{lstlisting}[language=Java, numbers=left, captionpos=b, 
label={JavaPackage.Editor.execute}, escapeinside={\%}{\%},
caption={Command execution}
]
package JavaPackages;
public class JavaPackagesEditor {
   ...
   public void execute(ModelCommand command) {
      String id = command.getId();
      if (id == null) {
         id = "obj" + activeCommands.size();
         command.setId(id);
      }
      String time = command.getTime();
      if (time == null) {
         time = getTime();
         command.setTime(time);
      }
      ModelCommand oldCommand = activeCommands.get(id);
      if (oldCommand != null && ! command.overwrites(oldCommand)) {
         return;
      }
      command.run(this);
      activeCommands.put(id, command);
   }   
   ...
\end{lstlisting}

When you want to execute a command you actually call \texttt{editor.execute(cmd)} on the 
responsible editor. Line~6 of Listing~\ref{JavaPackage.Editor.execute} first checks 
if your command has an \texttt{id}. If not, we create an \texttt{id} for you. This 
facilitates testing and iterative development. 
Alternatively, you may raise an exception. Next, Line~10 reads the time 
stamp of the command. If there is no time stamp yet, we assign the current time, 
cf. Line~13\footnote{The \texttt{getTime} method of our editor caches 
the time it returns. If you call it twice within the same millisecond, 
it will add an extra millisecond in order to avoid 
the same time stamp on multiple commands, cf. \cite{fulibServiceGeneratorGithub}}. 
If you serialize a command later on and send it e.g. to another editor, 
the command will already contain the original time stamp. 

Now Line~15 does a lookup in our hash table for \texttt{activeCommands}. 
If we already have an \texttt{oldCommand} we ask our new \texttt{command} 
whether it is going to \texttt{overwrite} the \texttt{oldCommand} (Line~16). 
Listing~\ref{ModelCommand.overwrites} shows the default implementation of 
method \texttt{overwrites}. Line~4 compares the time stamps and 
we do not overwrite if the current command is older than the \texttt{oldCommand}.
On equal time stamps (Line~6) it is probably the same command received twice
and there is no need to execute it again. However, for the unlikely case 
that two editors concurrently create different commands for the same \texttt{id} and with 
the same time stamp, we do a string compare of the yaml representation of our commands
and we do not execute the current command if the oldCommand is lexically later
or equal (Line~7 to 10). 

\begin{lstlisting}[language=Java, numbers=left, captionpos=b, 
label={ModelCommand.overwrites}, escapeinside={\%}{\%},
caption={Command execution}
]
public class ModelCommand {
   ...
   public boolean overwrites(ModelCommand oldCommand) {
      if (oldCommand.getTime().compareTo(time) > 0) {
         return false;
      } else if (oldCommand.getTime().equals(time)) {
         String oldYaml = Yaml.encode(oldCommand);
         String newYaml = Yaml.encode(this);
         if (oldYaml.compareTo(newYaml) >= 0) {
            return false;
         }
      }
      return true;
   }
   ...
\end{lstlisting}

If there is no \texttt{oldCommand} or if the new \texttt{command} overwrites 
the \texttt{oldCommand} our editor executes the new \texttt{command} 
in Line~19 of Listing~\ref{JavaPackage.Editor.execute}. 
Finally, the new \texttt{command} is added to the hash table of active commands
(Line~20). 

Thus, if you run (\texttt{HaveLeaf Editor serv 1.0}) in some editor 
at e.g. 13:36 o'clock and you 
run (\texttt{HaveLeaf Editor serv 1.1}) at e.g. 13:37 o'clock on the same editor, the second 
command will overwrite the first. If you run the two commands on two different editors, 
each editor will store its own command. If you do model synchronisation some time later, 
on the first editor the 13:37 command will overwrite the 13:36 command 
while on the other editor 
the 13:36 command will be ignored as that editor already has 
the 13:37 command for the same \texttt{id}. 
Eventually, both editors have the 13:37 command active and the \texttt{vTag} of the
\texttt{Editor} object will be \texttt{1.1}. 
Note, our model synchronisation scheme works also 
for two editors with different meta models. 

The default implementation of method \texttt{overwrites} 
shown in Listing~\ref{ModelCommand.overwrites} resolves editing or merge 
conflicts with a "last edit wins" strategy. While this works quite frequently, 
in some cases you may want another strategy for conflict resolution. For example 
in case of a seat reservation system you might want a "first edit wins" strategy. 
Or in our version number example you may want a "highest version wins" strategy. 
Therefore, each command may overwrite the inherited default implementation of 
method \texttt{overwrites} and implement its own strategy. If you do this, 
ensure that all editors (or models) use the same strategy during model 
synchronisation.

\section{Removing model objects} \label{sec.getOrCreate}

There is yet another design issue to be discussed. Due to our 
\texttt{getOrCreate} mechanism, removing an object from our model is 
quite tricky. To safely remove an object from our model we must get rid of
the command that created and initialized it directly (core object and green parts 
of our TGG rules, cf. Figure~\ref{fig:examplePatterns}). And we must get 
rid of all usages of this object as context (black rule part) in all other 
commands. Generally, we would require that for any object that is used as 
a required context (black rule part) by some command the corresponding 
set of \texttt{activeCommands}
shall contain another command that explicitly creates and initializes that 
object (core object and green rule part). 
Unfortunately, we deal with unreliable message brokers
and some commands may just not yet have arrived. But we already want to work 
with our model. Thus, it would be great to be able to 
distinguish between fully initialized 
\textit{model objects} and so-called \textit{object frames} that so far have been 
used as context objects, only. 
To achieve this, our editors deploy one 
\texttt{mapOfModelObjects} for explicit model objects and one 
\texttt{mapOfFrames} for context objects, 
cf. Listing~\ref{JavaPackagesEditor.maps} and \cite{fulibServiceGeneratorGithub}. 
(The \texttt{mapOfParsedObjects} will be discussed in Section~\ref{sec.parsing}.)

\begin{lstlisting}[language=Java, numbers=left, captionpos=b, 
label={JavaPackagesEditor.maps}, escapeinside={\%}{\%},
caption={Maps for model objects and object frames}
]
package JavaPackages;
...
public class JavaPackagesEditor {
   private Map<String, Object> mapOfModelObjects 
                 = new LinkedHashMap<>();
   private Map<String, Object> mapOfFrames 
                 = new LinkedHashMap<>();
   private Map<String, Object> mapOfParsedObjects 
                 = new LinkedHashMap<>();
   ...
}
\end{lstlisting}

When we need an object as context (black rule part) we call method 
\texttt{getObjectFrame} on the corresponding \texttt{editor}. 
Lines~7 to 10 of Listing~\ref{JavaPackagesEditor.getObjectFrame} will be 
discussed in Section~\ref{sec.parsing}.
Line~11 of Listing~\ref{JavaPackagesEditor.getObjectFrame} first tries 
to retrieve the desired object from the \texttt{mapOfModelObjects}. 
If that fails, Line~15 tries to retrieve the desired object from the 
\texttt{mapOfFrames}. 
If this still fails, Line~19 to 24 first use reflection in order to create 
the desired object and to initialize its \texttt{id} and then 
the object is added to the \texttt{mapOfFrames} and returned. 

\begin{lstlisting}[language=Java, numbers=left, captionpos=b, 
label={JavaPackagesEditor.getObjectFrame}, escapeinside={\%}{\%},
caption={getObjectFrame method}
]
package JavaPackages;
...
public class JavaPackagesEditor {
   ...
   public Object getObjectFrame(Class clazz, String id) {
      try {
         Object modelObject = mapOfParsedObjects.get(id);
         if (modelObject != null) {
            return modelObject;
         }
         modelObject = mapOfModelObjects.get(id);
         if (modelObject != null) {
            return modelObject;
         }
         modelObject = mapOfFrames.get(id);
         if (modelObject != null) {
            return modelObject;
         }
         modelObject = clazz.getConstructor().newInstance();
         Method setIdMethod 
                   = clazz.getMethod("setId", String.class);
         setIdMethod.invoke(modelObject, id);
         mapOfFrames.put(id, modelObject);
         return modelObject;
      } catch (Exception e) {
         throw new RuntimeException(e);
      }
   }
   ...
}
\end{lstlisting}

Similarly, when we want to create a model object explicitly 
(green rule parts) we call method \texttt{getOrCreate} on the 
corresponding editor, cf. Listing~\ref{JavaPackagesEditor.getOrCreate}. 
Again, Lines~6 to 10 will be discussed in 
Section~\ref{sec.parsing}.
Line~11 tries to retrieve the desired model object from 
our \texttt{mapOfModelObjects}. If that fails, Line~15 calls 
method \texttt{getObjectFrame} which either retrieves the 
desired object or creates it. Then Lines~16 and 17 promote the 
desired object into the \texttt{mapOfModelObjects} and Line~18
returns it. 

\begin{lstlisting}[language=Java, numbers=left, captionpos=b, 
label={JavaPackagesEditor.getOrCreate}, escapeinside={\%}{\%},
caption={getOrCreate method}
]
package JavaPackages;
...
public class JavaPackagesEditor {
   ...
   public Object getOrCreate(Class clazz, String id) {
      Object modelObject = mapOfParsedObjects.get(id);
      if (modelObject != null) {
         mapOfModelObjects.put(id, modelObject);
         return modelObject;
      }
      modelObject = mapOfModelObjects.get(id);
      if (modelObject != null) {
         return modelObject;
      }
      modelObject = getObjectFrame(clazz, id);
      mapOfFrames.remove(id);
      mapOfModelObjects.put(id, modelObject);
      return modelObject;
   }
   ...
}
\end{lstlisting}

In order to remove an object from our model we call 
method \texttt{removeModelObject} on the corresponding editor. 
Line~6 of Listing~\ref{JavaPackagesEditor.removeModelObject} 
tries to remove the object from our \texttt{mapOfModelObjects}. 
If that succeeds, Line~8 adds the removed object to our 
\texttt{mapOfFrames} (as it may still be used as context
by some other command). 

\begin{lstlisting}[language=Java, numbers=left, captionpos=b, 
label={JavaPackagesEditor.removeModelObject}, escapeinside={\%}{\%},
caption={removeModelObject method}
]
package JavaPackages;
...
public class JavaPackagesEditor {
   ...
   public Object removeModelObject(String id) {
      Object oldObject = mapOfModelObjects.remove(id);
      if (oldObject != null) {
         mapOfFrames.put(id, oldObject);
      }
      return mapOfFrames.get(id);
   }
   ...
}
\end{lstlisting}

Actually, to remove some object from our model, 
we need to (implement and) call 
the \texttt{remove} method of the responsible command in order to remove 
the complete increment and to leave our model in a consistent state. 
In addition, we have to remove 
the corresponding command from our \texttt{activeCommands} hash table. 
And we have to inform all other 
tools during subsequent model synchronisation. 
In addition, we have to be careful, in order 
to prevent the re-execution of the removed command when we receive it again (e.g. from 
another tool). Our implementation uses a special \texttt{RemoveCommand} to achieve this, 
cf. Listing~\ref{JavaPackage.RemoveCommand}. 

\begin{lstlisting}[language=Java, numbers=left, captionpos=b, 
label={JavaPackage.RemoveCommand}, escapeinside={\%}{\%},
caption={Command execution}
]
package JavaPackages;
...
public class RemoveCommand extends ModelCommand {
   public Object run(JavaPackagesEditor editor) {
      editor.removeModelObject(getId());
      ModelCommand oldCommand 
            = editor.getActiveCommands().get(getId());
      if (oldCommand != null) {
         oldCommand.remove(editor);
      }
      return null;
}
\end{lstlisting}

We may e.g. call (\texttt{RemoveCommand c}) on our \texttt{editor} and the 
\texttt{editor} may already have a (\texttt{HaveLeaf c sub 1.1}) command in
its set of \texttt{activeCommands}. Then the default "last edit wins" strategy of 
our \texttt{editor} will find that the \texttt{RemoveCommand} is later than 
the \texttt{HaveLeaf} command and it will call \texttt{run} on the 
\texttt{RemoveCommand}.  
Line~5 of Listing~\ref{JavaPackage.RemoveCommand} assumes per default 
that the command to be removed has created at least one model object with 
the same \texttt{id} as the command (and the \texttt{RemoveCommand}). Thus, 
Line~5 calls \texttt{removeModelObject} with that \texttt{id}. Therefore, 
simple commands that create and initialize only a single object do not 
even have to implement the \texttt{remove} method as the \texttt{RemoveCommand}
already does this job. If you do not want to rely on this default assumption, 
you may easily omit this line in your implementation. 

Line~7 of Listing~\ref{JavaPackage.RemoveCommand} retrieves the command to 
be removed from our \texttt{active\-Commands} and Line~9 calls its \texttt{remove}
method. Afterwards, the \texttt{execute} method of our \texttt{editor} 
replaces the \texttt{remove}d command with the \texttt{RemoveCommand} within our 
\texttt{activeCommands}, 
cf. Line~20 of Listing~\ref{JavaPackage.Editor.execute}. 
On model synchronization the \texttt{RemoveCommand} will be send to 
the other tool(s) and perform the same operation there. 

Note, we need to keep the \texttt{RemoveCommand} in our \texttt{activeCommands}
table until we are sure that all other tools (and all persistent copies of 
our \texttt{activeCommands} table have overwritten their copy of e.g. the 
\texttt{HaveLeaf} command. If we remove the \texttt{RemoveCommand} too early
and if we receive the overwritten \texttt{HaveLeaf} command thereafter, 
we would not notice that the \texttt{HaveLeaf} command has been removed but 
we would re-execute it.

\section{Parsing} \label{sec.parsing}

Sometimes, you may want to edit your model directly, 
cf. Line~13 to 24 of Listing~\ref{TestPackageToDoc.testManualChangesAndParsing}.
Then you need to parse your (modified) model in order to retrieve the 
set of commands that correspond to it and to synchronize your 
changes with other models. Parsing basically requires to split your model 
into increments where each increment corresponds to a certain command. 
In our approach, we follow the convention that each command creates one core
model object that gets the same \texttt{id} as the command. 
This core model object 
becomes the nucleus for each increment. 
Thus, in our approach 
the identification of model increments starts with these core 
objects and parsing just needs to collect the remaining parts of the increment.

\begin{lstlisting}[language=Java, numbers=left, captionpos=b, 
label={TestPackageToDoc.testManualChangesAndParsing}, escapeinside={\%}{\%},
caption={Command execution}
]
package JavaPackages;
...
public class TestPackageToDoc 
             implements PropertyChangeListener {
...
private Set changedObjects = new LinkedHashSet();
@Test
public void testManualChangesAndParsing() {
   JavaPackagesEditor javaPackagesEditor 
         = new JavaPackagesEditor();
   startSituation(javaPackagesEditor);
   registerModelObjectListener(javaPackagesEditor, this);
   JavaPackage com = new JavaPackage().setId("com");
   JavaPackage org = (JavaPackage) 
         javaPackagesEditor.getModelObject("org");
   com.withSubPackages(org);
   JavaPackage fulib = (JavaPackage) 
         javaPackagesEditor.getModelObject("fulib");
   fulib.setPPack(null);
   JavaClass command = new JavaClass()
         .setPack(fulib).setVTag("1.1").setId("Command");
   JavaClass editorClass = (JavaClass) 
         javaPackagesEditor.getModelObject("Editor");
   editorClass.setVTag("1.1");
   Set allObjects = Yaml.findAllObjects(com, fulib);
   javaPackagesEditor.parse(changedObjects);
   ...
}
...
}
\end{lstlisting}

In \cite{fritsche2020avoiding} the parsing of a \textit{HaveLeaf} rule requires 
that the \texttt{JavaPackage} that is reached via a \texttt{pack} link 
has already been parsed either by
a \texttt{HaveRoot} or by an \texttt{HaveSubUnit} rule. 
Similarly, the parsing of a \texttt{HaveSubUnit} rule requires that the 
\texttt{JavaPackage} attached to the corresponding sub \texttt{JavaPackage} 
via a \texttt{pPack - subPackages} link has already been parsed. 
Thus, in \cite{fritsche2020avoiding} parsing must start with the \texttt{HaveRoot}
rule and then you parse the sub \texttt{JavaPackages} of these root(s) and 
then the sub-sub packages until you reach the \texttt{JavaClass} leafs.
These parsing dependencies between general Triple Graph Grammar rules makes 
the parsing of Triple Graph Grammars (and especially incremental parsing) 
very complex.   

Compared to
general Triple Graph Grammars, parsing of Commutative Event Sourcing models 
is a piece of cake. 
As we have no rule dependencies, even incremental parsing 
becomes easy. Therefore, 
Line~12 of Listing~\ref{TestPackageToDoc.testManualChangesAndParsing} 
subscribes a property change listener to our object model. 
(See \cite{fulibServiceGeneratorGithub} for implementation details.) This 
property change listener collects all objects affected by the changes 
executed in Lines~13 to 24. This allows us to call the \texttt{parse}
method of our \texttt{editor} with just the set of \texttt{changedObjects}
in Line~26. Alternatively, Line~25 uses our object serialization mechanism
to collect all objects that belong to the modified model and we could 
use the set of \texttt{allObjects} within our \texttt{parse} call. 

Line~6 of Listing~\ref{Editor.parse} registers all objects 
that shall be parsed into our \texttt{mapOfParsedObjects} 
(cf. Listing~\ref{JavaPackagesEditor.maps}, see \cite{fulibServiceGeneratorGithub}
for details). For each object that shall be parsed, Line~10 calls method 
\texttt{findCommands}, which does the actual parsing. 
Method \texttt{findCommands} retrieves a set of command \texttt{prototypes}
(Line~26 of Listing~\ref{Editor.parse}). 
Then Line~28 calls the \texttt{parse} method of 
each command prototype. These parse methods analyse the current 
object, whether it is the nucleus of some increment that matches that 
command, cf. Listing~\ref{JavaPackage.HaveRoot.parse} and
Listing~\ref{JavaPackage.HaveSubUnit.parse}. If the command fits, 
its \texttt{parse} method returns a new copy of the corresponding command
with all command parameters assigned, properly. On success, 
Line~30 of Listing~\ref{Editor.parse} collects the parsed command in the 
set of \texttt{allCommands}. Once we have identified all 
commands that correspond to the objects to be parsed, we 
look for \texttt{oldCommand}s that will be overwritten by the 
execution of a new command, cf. Line~14 to 16 of Listing~\ref{Editor.parse}. 
If there is already an 
old command with the same parameters, we do not overwrite it in order to keep 
the old time stamp. If the new command is actually different
(or there is no old command) we execute it (Line~17) in order to 
update our set of \texttt{activeCommands}. (Note, executing a command
that has been parsed will edit the corresponding increment but this 
edit should just re-assign the values that have been found during parsing. 
Thus, if the \texttt{run} and the \texttt{parse} method work consistently, 
running the command does no harm (but may also be skipped).)

\begin{lstlisting}[language=Java, numbers=left, captionpos=b, 
label={Editor.parse}, escapeinside={\%}{\%},
caption={Command execution}
]
package JavaPackages;
...
public class JavaPackagesEditor {
...
public void parse(Collection allObjects) {
   registerParsedObjects(allObjects);
   ArrayList<ModelCommand> allCommandsFromParsing 
         = new ArrayList<>();
   for (Object object : allObjects) {
      findCommands(allCommandsFromParsing, object);
   }
   for (ModelCommand command : allCommandsFromParsing) {
      String id = command.getId();
      ModelCommand oldCommand = activeCommands.get(id);
      if (oldCommand == null || 
          ! equalsButTime(oldCommand, command)) {
         execute(command);
      }
   }
}

private ModelCommand findCommands(
      ArrayList<ModelCommand> allCommands, 
      Object object) {
   ArrayList<ModelCommand> prototypes 
         = haveCommandPrototypes();
   for (ModelCommand prototype : prototypes) {
      ModelCommand command = prototype.parse(object);
      if (command != null) {
         allCommands.add(command);
      }
   }
   return null;
}
...
}
\end{lstlisting}

There is still one little design issue to be discussed. 
In our example, Line~13 of Listing~\ref{TestPackageToDoc.testManualChangesAndParsing}
creates the \texttt{JavaPackage} \texttt{com}, directly. Our change listener 
will collect the new \texttt{com} object as soon as it is linked to the old 
model object \texttt{org} (Line~16). Thus, parsing will create a 
(\texttt{HaveRoot com}) command. When we execute this (\texttt{HaveRoot com}) 
command, its \texttt{run} method calls \texttt{getOrCreate} 
(Line~6 of Listing~\ref{JavaPackage.HaveRoot.run}) to retrieve 
the desired model object. Now, some parts of our (testing) program may still hold 
a reference to the directly created \texttt{com} object. Thus we want 
\texttt{getOrCreate} to retrieve this directly created \texttt{com} object 
and \texttt{getOrCreate} should not create a new model object. To achieve this, 
Line~6 of Listing~\ref{Editor.parse} registers all directly changed objects 
in our \texttt{mapOfParsedObjects} and method \texttt{getOrCreate} tries to 
retrieve the required object from there 
(Lines~6 to 10 of Listing~\ref{JavaPackagesEditor.getOrCreate}).
Method \texttt{getObjectFrame} work similarly 
(Listing~\ref{JavaPackagesEditor.getObjectFrame}).

\section{Conclusions}

To revisit the title of this paper, Commutative Event Sourcing may be 
considered as just a restricted variant of Triple Graph Grammars where 
the commands or rules are either overwriting or commutative. This frees 
Commutative Event Sourcing from handling dependencies between 
commands / rules. Thereby, model synchronisation, collaborative editing, 
and even incremental parsing is facilitated, considerably. 

This paper tries to provide sufficient details such that you may 
implement Commutative Event Sourcing for your tool(s), manually. 
You may adapt our concepts without relying on a special 
programming language, library, 
framework, or tool. You may also copy large parts of our example 
implementation from \cite{fulibServiceGeneratorGithub}. Actually, 
our editors are quite generic, only the set of command prototypes 
is model specific. \cite{fulibServiceGeneratorGithub} also 
provides a code generator for editors and the generic parts of 
commands, if you want to use that. \cite{fulibServiceGeneratorGithub}
even provides a simple interpreter for TGG like 
Commutative Event Sourcing patterns. However, using this rule 
interpreter requires a certain learning curve for writing these
patterns and pattern execution is hard to debug. Thus, beginners
may be better of by implementing their commands, manually.

Commutative event sourcing requires that commands / rules are 
either overwriting or commutative. This is quite a restriction 
compared to general Triple Graph Grammars. But it facilitates 
implementation. We like to compare this with the introduction 
of LALR(k) grammars \cite{deremer1982efficient} 
in compiler construction that reduced 
memory consumption within compilers compared to more general 
LR(k) grammars. LALR(k) grammars reduce the set of parseable 
languages, slightly, but acceptable. 

Our approach also makes extensive use of \texttt{id}s for 
(cross model) object reference. If you do not want 
\texttt{id} attributes in your model, your editor may use 
an additional \texttt{objectToId} map 
in order to store and retrieve object \texttt{id}s. 
We did so e.g. when using Commutative Event Sourcing for 
the solution of a model synchronisation case in the 
Transformation Tool Contest 2020 \cite{copeifulib} and 
\cite{fulibServiceGeneratorGithub}. 
If you use \texttt{id}s for cross model referencing, 
you want to assign these \texttt{id}s directly and 
you do not want e.g. a database system generating your 
\texttt{id}s. Directly edited \texttt{id}s are usually
considered an anti pattern. And yes, this is a design 
challenge for Commutative Event Sourcing. 

Commutative Event Sourcing considers two models to 
be equivalent if they contain the same objects with 
the same \texttt{id}s and same attribute values and the same 
\textit{set} of links. In case of a to-many association
we handle the neighbors as a set and not as a list, 
i.e. we ignore the order of the neighbors. 
If your commands are commutative and show up in any order
it is just hard to achieve a certain order in a list. 
This again is a design challenge, e.g. if you want to 
show a number of objects in the graphical user interface 
on two different tools and you want that both users see the
same order. (Well just sort them.) 
Neglecting the order of lists is also a challenge for 
models with a textual representation: you may not want 
an arbitrary order of your program statements (and sorting 
would not help).

However, we have used Commutative Event Sourcing with great 
success in \cite{copeifulib} and with some other model 
synchronisation problems e.g. for BPMN diagrams and a 
textual Workflow language \cite{fulibServiceGeneratorGithub}. 
We also used it in our \texttt{MicroServices} course in 
Winter Term 2019 / 2020 \cite{microservicesPlaylist}. 
According to our experiences, the Commutative Event Sourcing
approach to model synchronisation is quite easy to engineer
and works quite reliable.

\bibliography{references}

\begin{thebibliography}{10}
\providecommand{\url}[1]{\texttt{#1}}
\providecommand{\urlprefix}{URL }
\providecommand{\doi}[1]{https://doi.org/#1}

\bibitem{anjorin2017benchmarx}
Anjorin, A., Diskin, Z., Jouault, F., Ko, H.S., Leblebici, E., Westfechtel, B.:
  Benchmarx reloaded: A practical benchmark framework for bidirectional
  transformations. In: BX@ ETAPS. pp. 15--30 (2017)

\bibitem{copei2019mx}
Copei, S., Sälzer, M., Z{\"u}ndorf, A.: Multidirectional transformations for
  microservices. In: Proc. Dagstuhl Seminar. vol. 18491 (2019)

\bibitem{fulibServiceGeneratorGithub}
Copei, S., Z{\"u}ndorf, A.: Github example implementation.
  \url{https://github.com/fujaba/fulibServiceGenerator},
  \url{https://github.com/fujaba/fulibServiceGenerator/blob/master/test/src/test/java/
  javaPackagesToJavaDoc/TestPackageToDoc.java}, last viewed 31.08.2020

\bibitem{copeifulib}
Copei, S., Z{\"u}ndorf, A.: The fulib solution to the ttc 2020 migration case.
  arXiv preprint arXiv:2012.05231  (2020)

\bibitem{czarnecki2009bidirectional}
Czarnecki, K., Foster, J.N., Hu, Z., L{\"a}mmel, R., Sch{\"u}rr, A.,
  Terwilliger, J.F.: Bidirectional transformations: A cross-discipline
  perspective. In: International Conference on Theory and Practice of Model
  Transformations. pp. 260--283. Springer (2009)

\bibitem{deremer1982efficient}
DeRemer, F., Pennello, T.: Efficient computation of lalr (1) look-ahead sets.
  ACM Transactions on Programming Languages and Systems (TOPLAS)
  \textbf{4}(4),  615--649 (1982)

\bibitem{evans2004domain}
Evans, E.: Domain-driven design: tackling complexity in the heart of software.
  Addison-Wesley Professional (2004)

\bibitem{fritsche2020precedence}
Fritsche, L., Kosiol, J., M{\"o}ller, A., Sch{\"u}rr, A., Taentzer, G.: A
  precedence-driven approach for concurrent model synchronization scenarios
  using triple graph grammars. In: Proceedings of the 13th ACM SIGPLAN
  International Conference on Software Language Engineering. pp. 39--55 (2020)

\bibitem{fritsche2020avoiding}
Fritsche, L., Kosiol, J., Sch{\"u}rr, A., Taentzer, G.: Avoiding unnecessary
  information loss: Correct and efficient model synchronization based on triple
  graph grammars. arXiv preprint arXiv:2005.14510  (2020)

\bibitem{gamma1995design}
Gamma, E., Helm, R., Johnson, R., Vlissides, J.: Design patterns: Elements of
  reusable software architecture. Reading: Addison-Wesley  (1995)

\bibitem{qvt}
Group, O.M.: The qvt standard. \url{https://www.omg.org/spec/QVT/}, last viewed
  31.08.2020

\bibitem{hildebrandt2013survey}
Hildebrandt, S., Lambers, L., Giese, H., Rieke, J., Greenyer, J., Sch{\"a}fer,
  W., Lauder, M., Anjorin, A., Sch{\"u}rr, A.: A survey of triple graph grammar
  tools. Electronic Communications of the EASST  \textbf{57} (2013)

\bibitem{leblebici2014comparison}
Leblebici, E., Anjorin, A., Sch{\"u}rr, A., Hildebrandt, S., Rieke, J.,
  Greenyer, J.: A comparison of incremental triple graph grammar tools.
  Electronic Communications of the EASST  \textbf{67} (2014)

\bibitem{macedo2013implementing}
Macedo, N., Cunha, A.: Implementing qvt-r bidirectional model transformations
  using alloy. In: International Conference on Fundamental Approaches to
  Software Engineering. pp. 297--311. Springer (2013)

\bibitem{schneider2004coobra}
Schneider, C., Z{\"u}ndorf, A., Niere, J.: Coobra-a small step for development
  tools to collaborative environments. In: Workshop on Directions in Software
  Engineering Environments. Citeseer (2004)

\bibitem{schurr1994specification}
Sch{\"u}rr, A.: Specification of graph translators with triple graph grammars.
  In: International Workshop on Graph-Theoretic Concepts in Computer Science.
  pp. 151--163. Springer (1994)

\bibitem{steinberg2008emf}
Steinberg, D., Budinsky, F., Merks, E., Paternostro, M.: EMF: eclipse modeling
  framework. Pearson Education (2008)

\bibitem{stevens2010bidirectional}
Stevens, P.: Bidirectional model transformations in qvt: semantic issues and
  open questions. Software \& Systems Modeling  \textbf{9}(1), ~7 (2010)

\bibitem{microservicesPlaylist}
University, S.E.G.K.: Youtube playlist labor microservices, winter term 2019 /
  2020.
  \url{https://www.youtube.com/playlist?list=PLohPa1TMsVqrI0FaMTySbGr02JkQs5MhQ},
  last viewed 31.08.2020

\bibitem{vernon2013implementing}
Vernon, V.: Implementing domain-driven design. Addison-Wesley (2013)

\end{thebibliography}
\bibliographystyle{splncs04}

\end{document}